\documentclass[pre,aps,twocolumn,superscriptaddress,amsmath,showpacs,floatfix]{revtex4}
\usepackage{graphicx}
\begin{document}
\title{Duty-ratio of cooperative molecular motors}
\author{Nadiv Dharan}
\affiliation{Department of Biomedical Engineering,}
\author{Oded Farago}
\affiliation{Department of Biomedical Engineering,}
\affiliation{Ilse Katz Institute for Nanoscale Science and Technology, Ben Gurion University,
Be'er Sheva 84105, Israel}
\begin{abstract}
Molecular motors are found throughout the cells of the human body, and
have many different and important roles. These micro-machines move
along filament tracks, and have the ability to convert chemical energy
into mechanical work that powers cellular motility. Different types of
motors are characterized by different duty-ratios, which is the
fraction of time that a motor is attached to its filament. In the case
of myosin II - a non-processive molecular machine with a low duty
ratio - cooperativity between several motors is essential to induce
motion along its actin filament track. In this work we use statistical
mechanical tools to calculate the duty ratio of cooperative molecular
motors. The model suggests that the effective duty ratio of
non-processive motors that work in cooperation is lower than the duty
ratio of the individual motors. The origin of this effect is the
elastic tension that develops in the filament which is relieved when
motors detach from the track.
\end{abstract}
\maketitle

\section{Introduction}
\label{sec:intro}

Motor proteins are molecular machines that convert chemical energy
into mechanical work by ATP hydrolysis. They ``walk'' on the
microtubule and actin cytoskeleton and pull vesicles and organelles
across the cell \cite{Alberts:1994}. Motor proteins can be classified
into processive and non-processive motors. The former class includes
motors like kinesins, which travel a long distance along their
cytoskeleton track (microtubules) without detachment
\cite{kinesins}. In the latter class, we find motors like myosins that
make only a single step along their tracks (actin filaments) before
disconnection \cite{myosins}. While processive motors can move cargoes
by operating individually, non-processive motors need to work in
cooperation to generate substantial movement. Cooperative action of
myosin motors is implicated in a variety of cellular processes,
including the contraction of the contractile ring during cytokinesis
\cite{feierbach}, adaptation of mechanically activated transduction
channels in hair cells in the inner ear \cite{stauffer}, and muscle
contraction \cite{geeves}. Another important example of cooperative
motor dynamics is found in motility assays, where filaments glide over
a surface densely covered by motor proteins \cite{kron}.

One of the more interesting outcomes of cooperative action of
molecular motors is their ability to induce bidirectional
motion. Bidirectional motion is observed when a filament is subjected
to the action of two groups of motors that engage in a ``tug-of-war''
contest and exert forces in opposite directions \cite{tow}. The motor
party that exerts the larger force determines the instantaneous
direction of motion, which is reversed when the balance of forces
shifts from one group to the other. Earlier theoretical models
suggested that that the characteristic reversal time of the
bidirectional dynamics (i.e., the typical duration of the
unidirectional intervals of motion), $\tau_{\rm rev}$, grows
exponentially with the number of motors $N$ \cite{badoual}. In these
earlier models, the moving filament was treated as a rigid rod which
does not induce elastic coupling between the motors. A recent motility
assay of myosin II motors and actin filaments with alternating
polarities challenged this prediction. It was found that the
characteristic reversal times of the bidirectional motion in this
motility assay were macroscopically large, but practically independent
of the number of motors \cite{gilboa}. This observation has been
explained by a model that accounts for the elasticity of the actin
filament \cite{gilboa,gillo}. It has been shown that the motors
indirectly interact with each other via the tensile stress that they
generate in the elastic filament. The elasticity-mediated crosstalk
between the motors leads to a substantial increase in their unbinding
rates, making each motor effectively less processive and eliminating
the exponential growth of $\tau_{\rm rev}$ with $N$ \cite{farago}.

The reduction in the duty ratio (the fraction of time that each motor
spends attached to the filament track) of the cooperative motors can
be explained as follows: During bidirectional motion, the elastic
filament is subjected to a tug-of-war between motors that exert
opposite forces, which leads to large stress fluctuations along the
elastic filament. These stress fluctuations are the origin of the
elasticity-mediated crosstalk effect. Detailed analysis shows that the
typical elastic energy stored in the actin filament scales as
\cite{gilboa,gillo}
\begin{equation}
 \frac{E}{k_BT}= \alpha N n
\label{eq:1}
\end{equation}
where $N$ is the number of motors, $n$ is the number of {\em
connected}\/ motors, and $\alpha$ is a dimensionless parameter (which
is closely related to the parameter $\beta^*$, to be defined below in
eq.~\ref{eq:betastar}). For actin-myosin II systems, $\alpha$ is
rather small (for the motility assay described in ref.  \cite{gilboa},
~$\alpha\sim 2\times 10^{-3}$); but the energy released upon the
detachment of a single motor ($n\rightarrow n-1$): $ \Delta
E/k_BT=-\alpha N$, can be quite large if $N$ is large. This last
result implies that for large $N$, the transitions of motors between
the attached and detached states are influenced by both ATP-driven
(out-of-equilibrium) processes as well as by thermal (equilibrium)
excitations. The latter will be dominated by the changes in the
elastic energy of the actin (see eq.~\ref{eq:1}) and not by the energy
of the individual motors.

For the actin filament with alternating polarities, the
elasticity-mediated crosstalk is a cooperative effect that reduces the
degree of cooperativity between motors (decreases $\tau_{\rm rev}$) by
decreasing their duty ratio (i.e., their attachment probability). But
as discussed above, much of the strength of this effect is related to
the large stress fluctuations that develop in the elastic filament due
to the opposite forces applied by the antagonistic motors. In view of
this fact, it is fair to question the significance of this effect in
the more common situation where polar filaments move directionally
under the action of a single family of motor proteins. In this work we
analyze this problem and show that the elasticity-mediated crosstalk
effect in this system is indeed much smaller, but not entirely
negligible. Our analysis of this effect is presented in the following
section~\ref{sec:model}. The magnitude of this effect in acto-myosin
systems is estimated in section~\ref{sec:AMsys}.

\begin{figure}[t]
\begin{center}
\scalebox{0.5}{\centering \includegraphics{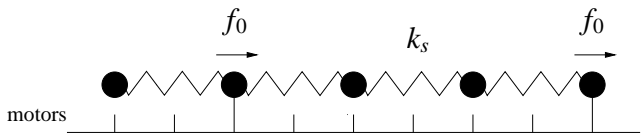}}
\end{center}
\vspace{-0.5cm}
\caption{The actin elastic filament is represented as a chain of nodes,
connected by identical springs with spring constant $k_s$. Each node
is either connected to a single myosin II motor - in which it
experiences a force of magnitude $f_0$, or disconnected - in which it
experiences no force.}
\label{fig:system}
\end{figure}

\section{Statistical-Mechanical Model}
\label{sec:model}

In what follows we model the elastic actin filament as a chain of $N$
equally spaced nodes connected by $N-1$ identical springs with spring
constant $k_s$ (see fig.~\ref{fig:system}). In the chain reference
frame, the $i$-th node is located at $x_i=(i-1)\Delta l$, where
$i=1,\ldots ,N$ and $\Delta l$ is the spacing between the nodes. For
brevity we set $\Delta l=1$. The chain lies on a ``bed'' of motors,
where each node may be either free and experience no pulling force
($f_i=0$), or attached to one motor in which case it is subjected to a
force of magnitude $f_i=f_0$. Different configurations of the system
are defined according to which nodes are connected to motors and which
are not. For a given configuration $j$, the elastic energy of the
chain is given by the sum of energies of the springs:
$E^{el}_j=\sum_{i=1}^{N-1}F^2_i/2k_s$, where $F_i$ is the force
applied on the $i$-th spring. The forces $F_i$ are calculated as
follows: We first calculate the mean force $\bar
{f}=\left(\sum_{i=1}^N f_i\right)/N$, and define the excess forces
acting on the nodes: $f_i^*=f_i-\bar{f}$. The force on the $i$-th
spring is then obtained by summing the excess forces applied on all
the monomers located on one side of the spring:
$F_i=-\sum_{l=1}^{i}f^*_l=\sum_{l=i+1}^Nf^*_l$.

\begin{figure}[b]
\vspace{0.5cm}
\begin{center}
\scalebox{0.35}{\centering \includegraphics{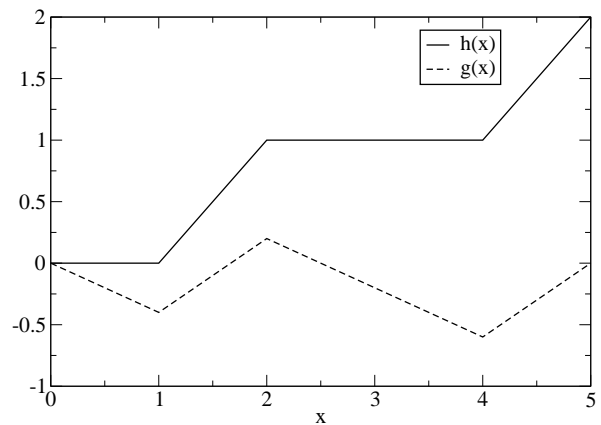}}
\end{center}
\vspace{-0.5cm}
\caption{The  functions  $h(x)$ and  $g(x)$  which  correspond to  the
configuration shown in fig.~\ref{fig:system}.}
\label{fig:hg}
\end{figure}

It is more convenient to analyze the problem using continuous
functions. Let us introduce the function $h(x)$ which, for
$x_i<x<x_{i+1}$, has slope $+1$ if the monomer at $x_i$ is connected
to a motor, and a slope 0, otherwise. Thus, $h(x)$ gives the total
force applied on the chain up to the point $x$, with $h(x=0)=0$ and
$h(x=N)=n$, where $n$ is the number of monomers connected to motors in
a given configuration. The solid line in fig.~\ref{fig:hg} shows the
function $h(x)$ corresponding to the configuration of 5 nodes depicted
in fig.~\ref{fig:system}. To calculate the elastic energy of a
configuration,we introduce the function $g(x)=h(x)-(n/N)x$, which is
depicted by the dashed line in fig.~\ref{fig:hg} and gives the total
{\em excess}\/ force accumulated up to $x$. The elastic energy can
then be expressed as:

\begin{equation}
\frac{E^{el}_j}{k_bT}=\beta^*\sum_{i=1}^{N-1}g^2(x_i)
\simeq\beta^*\int_{0}^{N}g^2(x)dx,
\label{eq:energy}
\end{equation}
where 
\begin{equation}
\beta^*=\frac{f_0^2}{2k_sk_BT}
\label{eq:betastar}
\end{equation}
is the ratio between the typical elastic energy of a spring
$f_0^2/2k_s$ and the thermal energy $k_BT$.

To determine the mean number of connected motors, one needs to
calculate the partition function
\begin{equation}
Z=\sum_{n=0}^N p^n(1-p)^{N-n}z_n,
\label{eq:Z}
\end{equation}
where $p$ is the attachment probability, i.e.~duty ratio, of a single
motor, and $z_n$ is the partition function of all the configurations
with exactly $n$ connected motors. The function $z_n$ can be
calculated by tracing over all the functions $g(x)$ corresponding to
configurations with $n$ connected motors. Mathematically, the
condition that exactly $n$ motors are connected can be expressed
through the following constraint on the function $h(x)$.
\begin{equation}
\lim_{\alpha \rightarrow \infty}\int_{0}^{N}\left |
\frac{dh}{dx} \right |^\alpha dx=n.
\label{eq:const1}
\end{equation}
\begin{widetext}
To allow an analytical solution, we approximate this constraint by
setting $\alpha=2$, in which case eq.~\ref{eq:const1} can be expressed
in terms of $g(x)=h(x)-\left (n/N \right )x$ as
\begin{equation}
\int_{0}^{N}\left ( \frac{dg}{dx} \right )^2dx=N
\frac{n}{N}\left ( 1-\frac{n}{N} \right )
\label{eq:const2}
\end{equation}
With eq.~\ref{eq:const2}, the partition function $z_n$ is given by
\begin{equation}
z_n=B\left(n,N\right)
\int{\cal D}\left[g(x)\right]\exp\left(-\beta^*\int_{0}^{N}g^2(x)dx\right)
\delta\left [ \int_{0}^{N}\left (
\frac{dg}{dx} \right )^2dx -N \frac{n}{N} \left (
1-\frac{n}{N} \right ) \right ] 
\label{eq:z_n1}
\end{equation}
where $\delta$ is Dirac's delta-function. The function
$B\left(n,N\right)$ is introduced in eq.~\ref{eq:z_n1} in order to
compensate for the error introduced by the approximated constraint
eq.~\ref{eq:const2}. We will determine this function through the
requirement that for $\beta^*=0$, i.e.~in the absence of elastic
crosstalk between the motors,
\begin{equation}
z_n|_{\beta^*=0}=\binom{N}{n}=\frac{N!}{n!(N-n)!},
\label{eq:z_n0}
\end{equation}
which is simply the number of ways to choose $n$ out of $N$ monomers.  

In order to calculate the partition function $z_n$, we use the Fourier
space representation of $\delta(x)$,
\begin{equation}
\delta(x-a)=\frac{1}{2\pi i}\int_{-i\infty}^{i\infty}e^{w(x-a)}dw,
\label{eq:Fdelta}
\end{equation}
and the Fourier series of $g(x)$,
\begin{equation}
g(x)=\sum_{k=-N/2}^{N/2-1}g_ke^{i\frac{2\pi}{N}kx}.
\label{eq:Fg}
\end{equation}
Substituting eqs.~\ref{eq:Fdelta} and ~\ref{eq:Fg} into
eq.~\ref{eq:z_n1} yields:
\begin{equation}
z_n=B\left(n,N\right)\frac{1}{2\pi
i}\int_{-i\infty}^{i\infty}dw\int{\cal D}\left[g_k\right]\exp\left [
wN\frac{n}{N}\left ( 1-\frac{n}{N} \right ) \right ] \times \exp\left
[ -\sum_k g_k^2\left ( \frac{8\pi^2}{N}k^2w+2N\beta^* \right ) \right
].
\label{eq:z_n2}
\end{equation}
Tracing over $g_k$ can be readily performed, giving
\begin{equation}
z_n=B\left(n,N\right)\frac{1}{2\pi i}\int_{-i\infty}^{i\infty}dw
\exp\left [
wN\frac{n}{N}\left ( 1-\frac{n}{N} \right ) \right ] \left\{ \prod_{k=0}^{N/2}
\frac{\pi}{2N\beta^*+8\pi^2k^2w/N} \right\}.
\label{eq:z_n3}
\end{equation}
The integral over $w$ can be evaluated using the method of steepest
descent, which yields:
\begin{equation}
z_n\simeq B\left(n,N\right)e^{G(w_0)},
\label{eq:steep}
\end{equation}
where, 
\begin{eqnarray}
G(w)&=&n\left ( 1-\frac{n}{N} \right )w-\sum_{k=0}^{N/2}\ln\left
[\frac{1}{\pi} \left ( \frac{8\pi^2}{N}k^2w+2N\beta^* \right )\right ]
\label{eq:G}
\\
&\simeq&N\left\{w\frac{n}{N}\left(1-\frac{n}{N}\right)-\frac{1}{2}\ln\left[
\frac{2e^{-2}N}{\pi}\left(\pi^2w+\beta^*\right)\right]-
\sqrt{\frac{\beta^*}{\pi^2w}}\tan^{-1}\left(\sqrt{\frac{\pi^2w}{\beta^*}}
\right)\right\}
\nonumber,
\end{eqnarray}
and $w_0$ satisfying 
\begin{equation} 
\frac{dG}{dw}{\Biggl
  |}_{w_0}=\frac{n}{N}\left(1-\frac{n}{N}\right)-\frac{1}{2w_0}+
\frac{1}{2}\sqrt{\frac{\beta^*}{\pi^2w_0^3}}\tan^{-1}\left(\sqrt{\frac{\pi^2w_0}{\beta^*}}\right)=0.
\label{eq:gtag}
\end{equation}
For $\beta^*\ll 1$, one gets
\begin{equation}
w_0\simeq \frac{N}{2n}\left(\frac{N}{N-n}\right)-
\sqrt{\frac{N}{8n}\left(\frac{N}{N-n}\right)\beta^*}.
\label{eq:w0samllb}
\end{equation}
From eqs.~\ref{eq:z_n0}, \ref{eq:steep}, \ref{eq:G}, and
\ref{eq:w0samllb}, one finds that 
\begin{equation}
B\left (n,N \right )=\binom{N}{n}e^{-G(w_0)}=
\binom{N}{n} \left ( \frac{\pi}{e^3} \right
)^{(N/2)}\left ( \frac{N^3}{n(N-n)} \right )^{N/2}
\label{eq:B}.
\end{equation}
Inserting eq.~\ref{eq:w0samllb} into eq.~\ref{eq:G}, and expanding
$G(w_0)$ in powers of $\sqrt{\beta*}$, yields
\begin{equation}
G(w_0)\simeq G(w_0)|_{\beta^*=0}
-\sqrt{\frac{n(N-n)}{2}\beta^*}.
\label{eq:Gbeta}
\end{equation}
Finally, for $\beta^*\ll1$, the partition function $z_n$ is obtained
by substituting eqs.~\ref{eq:B} and~\ref{eq:Gbeta} into
eq.~\ref{eq:steep}, which gives:
\begin{equation}
z_n\simeq\binom{N}{n}\exp\left(-\sqrt{\frac{n(N-n)}{2}\beta^*} \right ).
\label{eq:z_nfinal}
\end{equation}

In order to calculate the partition function $Z$, one needs to substitute
eq.~\ref{eq:z_nfinal} into into eq.~\ref{eq:Z}, which gives:
\begin{equation}
Z=\sum_{n=0}^N\binom{N}{n} p^n(1-p)^{N-n}
\exp\left(-\sqrt{\frac{n(N-n)}{2}\beta^*} \right )
\label{eq:Zfinal}.
\end{equation}
In the thermodynamic limit ($N\gg 1$), the sum in eq.~\ref{eq:Zfinal}
is dominated by one term which corresponds to the mean number of
attached motors $\langle n\rangle$. This term is given by
\begin{equation}
\langle n \rangle=
N\left[p-(1-2p)\sqrt{\frac{p(1-p)}{8}\beta^*}\right].
\label{eq:peff}
\end{equation}
From eq.~\ref{eq:peff} we identify the {\em effective} attachment
probability as
\begin{equation}
p_{\rm eff}\equiv \frac{\langle n \rangle}{N}
=p-(1-2p)\sqrt{\frac{p(1-p)}{8}\beta^*}
\label{eq:peff2}
\end{equation}
\end{widetext}
Notice that the second term on the right hand side of
eq.~\ref{eq:peff2} is anti-symmetric around $p=1/2$, and that for
$p<1/2$ ($p>1/2$), the effective attachment probability $p_{\rm eff}$
is smaller (larger) than $p$. This observation is directly related to
the fact the elasticity-mediated crosstalk effect is driven by the
tendency to reduce the force fluctuations along the elastic
filament. For $p<1/2$ ($p>1/2$) the force fluctuations are reduced by
the detachment (attachment) of motors, which brings the system closer
to the limiting case $p=0$ ($p=1$) where the force fluctuations
vanish.

\begin{figure}[t]
\begin{center}
\scalebox{0.35}{\centering \includegraphics{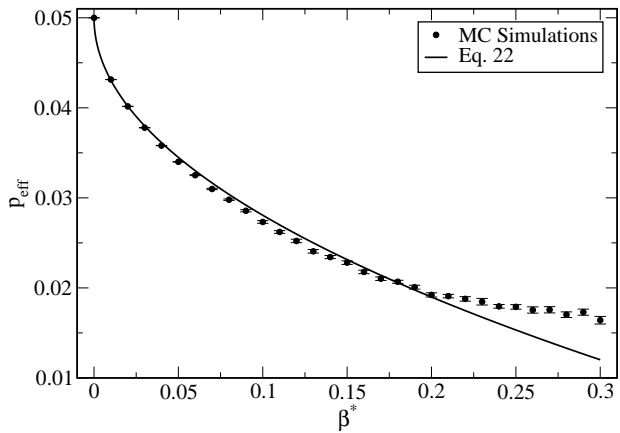}}
\end{center}
\vspace{-0.5cm}
\caption{The effective attachment probability as a function of the
dimensionless parameter $\beta^*$. The circles denote the results of
the MC simulations, while the solid line depicts the analytical
approximation for $\beta^*\ll 1$, eq:~\ref{eq:peff2}.}
\label{fig:MC}
\end{figure}

To test the validity and range of applicability of eq.~\ref{eq:peff2},
we conducted Monte Carlo (MC) simulations of elastic chains of
$N=1000$ monomers with $p=0.05$, which is the typical duty ratio of
myosin II motors \cite{finer}. Systems corresponding to different
values of $\beta^*$ were simulated using the parallel tempering
method. The simulations include two types of elementary moves (which
are attempted with equal probability) - one in which the state
(connected/disconnected) of a randomly chosen node changes, and the
other in which two randomly chosen nodes with opposite states change
their states simultaneously. For each move attempt, the model energy
of the chain is recalculated, and the move is accepted/rejected
according to the conventional Metropolis criterion. Our MC results
are summarized in fig.~\ref{fig:MC}. For $\beta^*<0.2$, we find an
excellent agreement between our computational results and
eq.~\ref{eq:peff2} (which has been derived for $\beta^*\ll 1$). Notice
that eq.~\ref{eq:peff2} does not include {\em any}\/ fitting
parameters. For larger values of $\beta^*$, eq.~\ref{eq:peff2}
overestimates the decrease in $p_{\rm eff}$.

\section{Actin-Myosin II systems}
\label{sec:AMsys}

Eq.~\ref{eq:betastar} relates the dimensionless parameter $\beta^*$
to three physical parameters of the system - the temperature $T$, the
typical force exerted by the motors $f_0$, and the effective spring
constant of the actin filaments segment between between two binding
sites of the motors, $k_s$. The latter parameter can be further expressed as
\begin{equation}
k_s=\frac{YA}{l},
\label{eq:ks}
\end{equation}
where $Y$ is Young's modulus of the actin, $A$ is the cross-sectional
area of an actin filament and $l$ is the distance between binding
sites. For myosin II motors, forces in the range of $f_0=5-10$ pN have
been measured experimentally \cite{molloyA,finer}. The actin-cross
sectional area (including the tropomyosin wrapped around the actin
helix) is $A=23\ {\rm nm}^2$, and the Young's modulus of the
actin-tropomyosin filament is $Y=2.8\ {\rm GPa}$
\cite{higuchi,kojima}. The value of $l$ is somewhat more difficult to
assess. One possibility is that $l\simeq 5.5$ nm, which is simply the
size of the G-actin monomers, each of which includes one binding site
for myosin motors \cite{holmes}.  Another possible value is related to
the double helical structure of F-actin and the fact that it completes
half a twist about every 7 monomers, i.e.~every $\simeq38$ nm
\cite{hooper}. Since the binding sites follow a twisted spatial path
along the double helix, many of them remain spatially unavailable to
the motors. In the motility assay, the motors are located underneath
the F-actin, and the distance between the binding sites along the line
of contact with the bed of motors is $l=38$ nm. A third choice for
$l$, which may be more relevant to skeletal muscles, is $l=14$
nm. This value is derived from the fact that within the sarcomere (the
basic contractile unit of the muscle) an average number of three thick
myosin filaments surround one thin actin filament. The separation
between collinear motor heads along the thick filament is $\sim 43$ nm
\cite{huxley,wakabayashi,daniel}, and because the actin is surrounded
by three thick filaments, the distance between the motors {\em along
the actin}\/ is $l=43/3\simeq 14$ nm.

For the duty ratio $p$, the range of experimental values is scattered
and varies from $p=0.01-0.02$ \cite{Howard:2001} to $p=0.05$
\cite{finer}. The uncertainty may be partially related to the
elasticity cross-talk effect discussed here. It also stems from the
fact that $p$ also depends on the distance and orientation of the
motor head (both in space and time) with respect to the associated
binding site. These may vary between the motors which, hence, should
have different attachment probabilities \cite{footnote}.

Using the above mentioned values of system parameters, one finds that
for acto-myosin systems the parameter $\beta^*$ lies within the range
of $5\times 10^{-4}\lesssim\beta^*\lesssim5\times 10^{-3}$. Setting
the duty ratio of myosin II to $p=0.05$ (as used in the MC
simulations) we find that for this range of $\beta^*$, the {\em
effective} duty ratio is slightly lower than $p$ and lies within the
range of $p_{\rm eff}=0.97p$ (for $\beta^*=5\times 10^{-4}$) and
$p_{\rm eff}=0.90p$ (for $\beta^*=5\times 10^{-3}$). Using a lower
estimate for the duty ratio $p=0.02$ \cite{Howard:2001}, we find that
the effective duty ratio drops to $0.83p\lesssim p_{\rm eff}\lesssim
0.95p$ for the same range of $\beta^*$.

As stated earlier, the cooperative action of myosin motors
``compensates'' for the non-processive character of the individual
motors. The force generated by a group of $N$ motors is $\langle F
\rangle=Np_{\rm eff}f_0$. Which force $f_0$ maximizes the effective
force per motor $f_{\rm eff}=\langle F \rangle/N$ and, hence, the
force production of the collectively working motors? From
eq.~\ref{eq:betastar} and~\ref{eq:peff2} we find that the maximum
value of $f_{\rm eff}$ is achieved when the force of the individual
motors is
\begin{equation}
f_0^{\rm max}=\frac{2p}{1-2p}\sqrt{\frac{k_sk_BT}{p(1-p)}}
\label{eq:f0max}
\end{equation}
Setting the values of the system parameters as above ($Y=2.8\ {\rm
GPa}$, $A=23\ {\rm nm}^2$, $l=38\ {\rm nm}$) and taking $p=0.02$ as
the duty ratio of a single motor, we find that $f_0^{\rm max}\simeq
25\ {\rm pN}$, for which $f_{\rm eff}\simeq0.25\ {\rm pN}$. For forces
in the range of $f_0\simeq 5-10\ {\rm pN}$, which are typically
measured for myosin II motors \cite{finer}, the effective mean force
per motor is $f_{\rm eff}\simeq0.1-0.15\ {\rm pN}$, which is about half
of the optimal effective force $f_{\rm eff}\left( f_0^{\rm
max}\right)$. We, thus, conclude that mysoin II motors work quite
close to conditions that maximize their cooperative force generation.

\section{Discussion}

There has recently been a considerable interest in the collective
behavior of molecular motors, especially in relation to cooperative
dynamics of cytoskeletal filaments in motility assays. Most studies
have focused on the bidirectional motion arising when the filament is
driven by two groups of antagonistic motors, or in the case when one
motor party works against an external force. The present work is
motivated by our recent studies of bidirectional motion in acto-myosin
motility assays which demonstrated that the duty ratio of the motors
(and, hence, the level of cooperativity between them) is reduced by
the elasticity of the actin backbone. In this paper we extended our
studies to motility assays in which similar motors act on a polar
actin filament without a counter external force. To single out the
filament elasticity crosstalk from other possible collective effects
(e.g., those associated with non-equilibrium ATP-assisted processes
and with the elasticity of the motors themselves
\cite{kruse,guerin,banerjee}), we neglect motor-to-motor variations
and use a model in which the motors are characterized by two mean
quantities: their attachment probability to a rigid (non-elastic)
filament $p$, and the mean applied pulling force $f_0$. We expect this
mean field description of the motors to hold when the number of motors
$N$ becomes large. We calculate the attachment probability to the
elastic filament $p_{\rm eff}<p$ from the partition function
associated with the filament elastic energy. The elastic energy of
filament can be treated as an equilibrium degree of freedom (of a
system which is inherently out-of-equilibrium) because the mechanical
response of the filament to the attachment/detachment of motors occurs
on time scales which are far shorter than the typical attachment time
of the motors (see discussion in ref.~\cite{farago}. The assumption
that the actin filament is in mechanical equilibrium is also made in
theoretical studies of intracellular cargo transport \cite{tow}
and muscle constraction \cite{duke}) Our calculation shows that
$p_{\rm eff}$ is only slightly smaller than $p$. This result is very
different from our previous findings for motility assays of
bidirectional motion, where the elasticity-mediated crosstalk effect
is substantial. Finally, we note that although in both this and
previous studies we found that $p_{\rm eff}<p$ \cite{gillo,gur}, the
opposite relation cannot be excluded under certain conditions (e.g.,
when the filament experiences external forces as in single molecule
experiments or muscle contraction). We intend to address this opposite
scenario in a future publication, in which we investigate the
variations in both $p$ (the ATP-hydrolysis related attachment
probability) and $p_{\rm eff}$ (which also includes the contribution
due to the elastic crosstalk effect) with the muscle contraction
velocity.

We thank Anne Bernheim and Sefi Givli for useful discussions.
\appendix

\section{The elastic energy}

\begin{figure}[h]
\begin{center}
\scalebox{0.55}{\centering \includegraphics{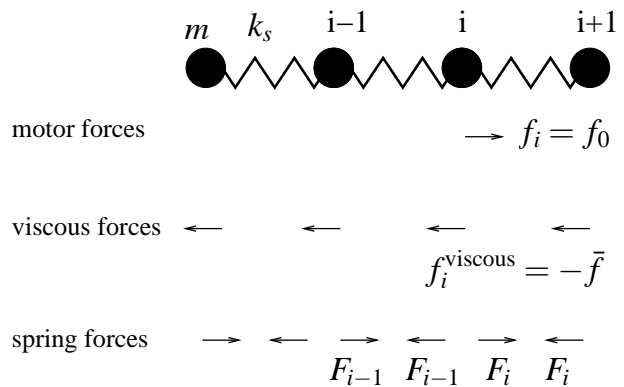}}
\end{center}
\vspace{-0.5cm}
\caption{The actin is modeled as a chain of masses and springs that
moves in a highly viscous medium. Three types of forces are present in
the system: the motor forces, a uniform viscous drag, and the spring
forces. The motion is highly overdamped (zero acceleration) and, thus,
the net force on each mass vanishes.}
\label{fig:system2}
\end{figure}

We model the actin filament as a linear chain of identical particles
of mass $m$ connected by elastic (massless) springs with spring
constant $k_s$ (see fig.~\ref{fig:system2}). Three types of forces are
exerted on the particles: (i) the motor forces $f^{\rm motor}_i$, (ii)
the spring forces $F_i$, and (iii) friction drag forces $f^{\rm
drag}_i$. Because the motion is highly overdamped, the total
instantaneous force on each mass vanishes. For the $i$-th mass, the
equation of is
\begin{equation} 
f^{\rm motor}_i-f^{\rm drag}_i+F_i-F_{i-1}=0,
\label{eq:overdamped}
\end{equation}
with $F_0=F_{N}=0$.  The drag force includes two contributions - one
is due to motor friction (MF), $f^{\rm MF}_i$, and the other due to
friction with the viscous medium, $f^{\rm viscous}_i$. The motor
friction forces act only on the particles which are connected to the
motors. Therefore, by redefining the motor forces $f_i=f^{\rm
motor}_i-f^{\rm MF}_i$, we can rewrite eq.~\ref{eq:overdamped} as
\begin{equation} 
f_i-f^{\rm viscous}_i+F_i-F_{i-1}=0.
\label{eq:overdamped2}
\end{equation}
For the chain's center of mass (CoM), the equation of motion reads
\begin{equation} 
F^{\rm CoM}=\sum_{i=1}^N f_i- \sum_{i=1}^N f^{\rm viscous}_i=0;
\label{eq:overdamped3}
\end{equation}
but the viscous drag force is distributed uniformly along the chain
(i.e., it is equal for all masses) and, therefore,
eq.~\ref{eq:overdamped3} gives
\begin{equation} 
f^{\rm viscous}_i=\frac{\sum_{i=1}^N f_i}{N}=\bar{f}.
\label{eq:overdamped4}
\end{equation}
Using this last result, eq.~\ref{eq:overdamped2} reads
\begin{equation} 
F_i-F_{i-1}=-f_i^*,
\label{eq:overdamped5}
\end{equation}
where $f_i^*$ is the excess force. Since $F_0=0$, we find that
$F_1=-f_1^*$. Then, $F_2=F_1-f_2^*=-f_1^*-f_2^*$; and, in general,
 \begin{equation} 
F_i=-\sum_{i=1}^N f_i^*.
\label{eq:final}
\end{equation}  

If the motion is only partially overdamped (including in the limit of
no friction), all the masses move together at the same
acceleration. One can repeat the above calculation and show that
eq.~\ref{eq:final} remains valid.



\newpage
\begin{thebibliography}{99}

\bibitem{Alberts:1994} B. Alberts, D. Bray, J. Lewis, M. Raff,
K. Roberts, and J. D. Watson, {\it Molecular Biology of the Cell}\/
(Garland, New York, 1994).

\bibitem{kinesins} H. Miki, Y. Okada, and N. Hirokawa, Trends. Cell
Biol.{\bf 15}, 467 (2005).

\bibitem{myosins} L. M. Coluccia (ed.), {\it Myosins: A
Superfamily of Moleculare Motors}\/ (Springer, Dordrecht, 2008).

\bibitem{feierbach} B. Feierbach and F. Chang,
Curr. Opin. Microbiol. {\bf 4}, 713 (2001).

\bibitem{stauffer} E. A. Stauffer {\em et al.}\/, Neuron {\bf 47}, 541
  (2005).

\bibitem{geeves} M. A. Geeves and K. C. Holmes,
Annu. Rev. Biochem. {\bf 68}, 687 (1999).

\bibitem{kron} S. J. Kron and J. A. Spudich,
Proc. Natl. Acad. Sci. USA {\bf 83}, 6272 (1986).

\bibitem{tow} M. J. I. M\"{u}ller, S. Klumpp, and R. Lipowsky,
Proc. Natl. Acad. Sci. USA {\bf 105}, 4609 (2008).

\bibitem{badoual} M. Badoual, F. J\"{u}licher, and J. Prost,
Proc. Natl. Acad. Sci. USA {\bf 99}, 6696 (2002).

\bibitem{gilboa} B. Gilboa, D. Gillo, O. Farago, and
A. Bernheim-Groswasser, Soft Matter {\bf 5}, 2223 (2009).

\bibitem{gillo} D. Gillo, B. Gur, A. Bernheim-Groswasser, and
O. Farago, Phys. Rev. E {\bf 80}, 021929 (2009).

\bibitem{farago} O. Farago and A. Bernheim-Groswasser, Soft Matter
{\bf 7}, 3066 (2011)..

\bibitem{finer} J. T. Finer, R. M. Simmons, and J. A. Spudich,  
Nature {\bf 368}, 113 (1994). 

\bibitem{molloyA} J. E. Molloy, J. E. Burns, J. Kendrick-Jones, R. T. Tregear, and D. C. S. White, Nature {\bf 378}, 209 (1995).

\bibitem{higuchi} H. Higuchi, T. Yanagida, and Y. E. Goldman, 
Biophys. J. {\bf 69}, 1000 (1995).

\bibitem{kojima} H. Kojima, A. Ishijima, and T. Yanagida,
Proc. Natl. Acad. Sci. USA {\bf 91}, 12962 (1994).

\bibitem{holmes} K. C. Holmes, D. Popp, W. Gebhard, and W. Kabsch, Nature {\bf 347}, 44 (1990).

\bibitem{hooper} S. L. Hooper, K. H. Hobbs, and J. B. Thuma,  
Prog. Neurobiol. {\bf 86}, 72 (2008).

\bibitem{huxley} H. E. Huxley, A. Stewart, H. Sosa, and T. Irving, Biophys. J. {\bf 67},
2411 (1994).

\bibitem{wakabayashi} K. Wakabayashi, Y. Sugimoto, H. Tanaka, Y. Ueno,
Y. Takezawa, and Y. Amemiya, Biophys. J. {\bf 67}, 2422 (1994).

\bibitem{daniel} T. L. Daniel, A. C. Trimble, and P. B. Chase,
Biophys. J. {\bf 74}, 1611 (1998).

\bibitem{Howard:2001} J. Howard, {\it Mechanics of Motor Proteins
and the Cytoskeleton}\/ (Sinauer, Sunderland MA, 2001).

\bibitem{footnote} In our statistical-mechanical analysis, $p$
represents the typical duty ratio of individual myosin II motors,
which are all assumed to have the same fixed $p$. It is possible that
accounting for the variations in $p$ in the model will result in a
slight increase in the magnitude of the elasticity cross talk effect.

\bibitem{kruse} K. Kruse and D. Riveline, Curr. Top. Dev. Biol. {\bf
95}, 67 (20110.

\bibitem{guerin} T. Gu\'{e}rin, J. Prost, P. Martin,
and. J.-F. Joanny, Curr. Opin. Ceel Biol. {\bf 22}, 14 (2010).

\bibitem{banerjee} S. Banerjee, M. C. Marchetti, and
K. M\"{u}ller-Nedebock, Phys. Rev. E. {\bf 84}, 011914 (2011).

\bibitem{duke} T. A. J. Duke, Proc. Nat. Acad. Sci. USA {\bf 96}, 2770
(1999).


\bibitem{gur} B. Gur and O. Farago, Phys. Rev. Lett. {\bf 104}, 238101
(2010).


\end{thebibliography}
\end{document}